\documentclass[twocolumn,showpacs,preprintnumbers,amsmath,amssymb]{revtex4}

\usepackage{graphicx}
\usepackage{dcolumn}
\usepackage{bm}

\begin{document}

\preprint{IUHET-500}

\title{Combined flavor symmetry violation and lepton number violation in 
neutrino physics}

\author{Micheal S. Berger}
 \email{berger@indiana.edu}
\author{Samuel Santana}%
 \email{sasantan@indiana.edu}
\affiliation{%
Physics Department, Indiana University, Bloomington, IN 47405, USA}

\date{\today}

\begin{abstract}
Heavy singlet neutrinos admit Majorana masses which are not possible for 
the Standard Model particles. This suggest new possibilities for generating
the masses and mixing angles of light neutrinos. We present a model of 
neutrino physics which combines the source of lepton number violation with 
the flavor symmetry responsible for the hierarchy in the charged 
lepton and quark sector. This is accomplished by giving the scalar 
field effecting the lepton number violation a nonzero charge under the 
horizontal flavor symmetry. We find an economical model which 
is consistent with the measured values of the atmospheric and solar neutrino
mass-squares and mixing angles.
\end{abstract}

\maketitle

\def\al{\alpha}
\def\be{\beta}
\def\ga{\gamma}
\def\de{\delta}
\def\ep{\epsilon}
\def\ve{\varepsilon}
\def\ze{\zeta}
\def\et{\eta}
\def\th{\theta}
\def\vt{\vartheta}
\def\io{\iota}
\def\ka{\kappa}
\def\la{\lambda}
\def\vpi{\varpi}
\def\rh{\rho}
\def\vr{\varrho}
\def\si{\sigma}
\def\vs{\varsigma}
\def\ta{\tau}
\def\up{\upsilon}
\def\ph{\phi}
\def\vp{\varphi}
\def\ch{\chi}
\def\ps{\psi}
\def\om{\omega}
\def\Ga{\Gamma}
\def\De{\Delta}
\def\Th{\Theta}
\def\La{\Lambda}
\def\Si{\Sigma}
\def\Up{\Upsilon}
\def\Ph{\Phi}
\def\Ps{\Psi}
\def\Om{\Omega}
\def\mn{{\mu\nu}}
\def\cD{{\cal D}}
\def\cF{{\cal F}}
\def\cL{{\cal L}}
\def\cS{{\cal S}}
\def\fr#1#2{{{#1} \over {#2}}}
\def\frac#1#2{\textstyle{{{#1} \over {#2}}}}
\def\pt#1{\phantom{#1}}
\def\prt{\partial}
\def\vev#1{\langle {#1}\rangle}
\def\ket#1{|{#1}\rangle}
\def\bra#1{\langle{#1}|}
\def\amp#1#2{\langle {#1}|{#2} \rangle}
\def\half{{\textstyle{1\over 2}}}
\def\lsim{\mathrel{\rlap{\lower4pt\hbox{\hskip1pt$\sim$}}
    \raise1pt\hbox{$<$}}}
\def\gsim{\mathrel{\rlap{\lower4pt\hbox{\hskip1pt$\sim$}}
    \raise1pt\hbox{$>$}}}
\def\ol#1{\overline{#1}}
\def\Re{\hbox{Re}\,}
\def\Im{\hbox{Im}\,}
\def\etal {{\it et al.}}
\def\slash#1{\not\hbox{\hskip -2pt}{#1}}

\section{Introduction}

The measurement of neutrino oscillations has created opportunities and 
challenges for model builders. The additional information regarding the 
masses and mixing angles in the neutrino sector provide valuable targets 
for models which aspire to explain the experimental data. 
The fact that the data has a combination of what appear to be 
large and small mixing angles as well as what might be at least a 
moderate hierarchy in neutrino masses is not what usually results in a generic 
way from the most simple models.

Nir and Shadmi\cite{Nir:2004my,Nir:2004pw} made an interesting observation 
regarding the unique characteristics of the neutrino sector where the 
right-handed neutrino is a Standard Model gauge singlet. In their framework
new possibilities exist for generating mass terms because there is a new
scale associated with the fact that the heavy neutrinos can have a 
Majorana mass
that violates lepton number.
They imagine a symmetry 
\begin{eqnarray}
G_{SM}\times U(1)_H\times U(1)_L\;,
\label{symmetry}
\end{eqnarray}
where $G_{SM}$ is the Standard Model gauge group, $U(1)_H$ is the usual
horizontal (or flavor) symmetry of Froggatt-Nielsen\cite{Froggatt:1978nt}
models, and $U(1)_L$ is 
lepton number. They propose (in the usual way) a scalar field $S_H$ which 
carries charge $-1$ under $U(1)_H$ which generates the mass and mixing angle
hierarchies in the charged lepton and quark sectors of the Standard Model.
In addition there is a pair of scalar fields $S_L$ and $\bar{S}_L$ that are 
singlets under $G_{SM}\times U(1)_H$ and have lepton number $-2$ and 
$+2$. The symmetry breaking involving these fields effects lepton 
number violation and allows for nonzero neutrino masses.
They looked at a few examples in the case of two light neutrino generations to 
illustrate their point in the setting of the Froggatt-Nielsen framework. 
In particular they showed that neutrino anarchy can 
result even when the horizontal flavor charges of the charged lepton families
are different for each generation. 

Related ideas have been explored by Barr and Kyae\cite{Barr:2004dr,Barr:2005ey}
in grand unified models. In their papers the new interactions are introduced
into fully specified grand unified theories, but the source of the new
contributions to the mass matrices comes from integrating out of the theory 
heavy vectorlike fermions like those in a Froggatt-Nielsen models.

\section{Specification of the Model}

In this paper we provide a realistic explanation for the neutrino 
oscillation data which can be obtained in a very economical way. The 
fundamental new feature we add to the models examined in 
Ref.~\cite{Nir:2004my,Nir:2004pw} is to give the lepton number violating scalar
field a horizontal charge. We refer to this as ``combining'' the lepton 
number violation with the flavor symmetry violation.
The model we present here falls into a class that 
implements an $L_e-L_\mu -L_\tau$ symmetry. This results in a 
pseudo-Dirac form for the 
mixing between the electron neutrino and the maximally mixed state from 
the muon- and 
tau-neutrino sector. It has been 
noted\cite{Petcov:1982ya,Barbieri:1998mq,Goh:2002nk,Frampton:2004ud,Petcov:2004rk} 
that a 
light neutrino mass matrix of the form
\begin{eqnarray}
m\left (\begin{array}{ccc}
0 & 1 & 1 \\
1 & 0 & 0 \\
1 & 0 & 0 
\end{array}\right )\;,
\end{eqnarray}
has two large (in fact maximal) mixing angles, a contribution 
$U_{e3}=0$ and an inverted
hierarchy. It also has $\Delta m^2=0$ and $m^2$ which is very roughly the 
pattern observed in neutrino oscillations.
Our model achieves approximately this form (absent the extra $\mu$-$\tau$
symmetry that guarantees the exact bimaximal form) but implements 
perturbations, consistent with Eq.~(\ref{symmetry}),
that are needed to be in agreement with the solar
and atmospheric neutrino data.

In Ref.~\cite{Nir:2004my} examples were provided that involved two light
neutrino generations and that exhibited the variety of behavior that can arise
in that case.
In this section we introduce a full three-generation model. 
The importance of neutrinos
being Majorana plays a crucial role in the properties of the light 
neutrino mass and mixing angles. The model implements an accidental 
$L_e-L_\mu -L_\tau$ 
symmetry\cite{Petcov:1982ya,Barbieri:1998mq,Goh:2002nk,Frampton:2004ud,Petcov:2004rk} 
that is broken perturbatively by the interactions of the Froggatt-Nielsen 
fields whose vevs break the horizontal symmetry $U(1)_H$.

Following Nir and Shadmi we assume there is a scalar field $S_H$ that is a 
singlet of $G_{SM}\times U(1)_L$ and carries charge $-1$ under $U(1)_H$. This
field is the usual Froggatt-Nielsen scalar whose vacuum expectation value (vev)
generates the hierarchy in the charged lepton and quark sectors. We will 
also assume there is another scalar field $S_H^\prime$ that is a singlet
under $G_{SM}\times U(1)_L$ and carries charge $+1$ under $U(1)_H$. The fields
$S_H$ and $S_H^\prime$ admit interactions that are incompatible
with the $L_e-L_\mu -L_\tau$ symmetry and as a consequence generate 
the necessary perturbations that make the model phenomenologically viable. 
The lepton 
number symmetry $U(1)_L$ is broken by two scalar fields $S_{HL}$ and 
$\bar{S}_{HL}$ which carry charges $-2$ and $+2$ respectively, under $U(1)_L$. 
We also assign these two fields charges $-1$ and $+1$ respectively, under 
$U(1)_H$. It is in this last respect that our model differs from the explicit
examples of Nir and Shadmi who chose to make the lepton number violating 
scalars (called $S_L$ and $\bar{S}_L$) singlets under the horizontal 
symmetry $U(1)_H$. For this reason we have chosen to denote this lepton number 
violating field $S_{HL}$ (rather than $S_L$).

The scales in the model are: 

1) the electroweak breaking scale determined by 
the vevs of Higgs doublets, $\langle \phi_{u,d}\rangle$;

2) $M_{HL}\equiv \langle S_{HL}\rangle \sim \langle \bar{S}_{HL}\rangle$, the 
lepton number breaking scale

3) $M_H\equiv \langle S_H\rangle \sim \langle S_H^\prime\rangle$, 
the horizontal symmetry breaking scale;

4) $M_F$, the mass scale of Froggatt-Nielsen vector-like quarks and leptons

The Froggatt-Nielsen parameter 
\begin{eqnarray}
\lambda_H\equiv {{\langle S_H\rangle}\over M_F}={M_H\over M_F}\;,
\end{eqnarray} 
is small and is often associated with the value of Cabibbo angle ($\sim 0.2$)
since it is used to generate the mass and mixing angle hierarchies in the quark
and charged lepton sectors of the Standard Model.
The lepton number breaking introduces another parameter 
\begin{eqnarray}
\lambda _L^2\equiv {{\langle S_H \rangle ^2}\over {\langle S_{HL}\rangle
\langle \bar{S}_{HL}\rangle}}={M_H^2\over M_{HL}^2}\;,
\label{lambdal}
\end{eqnarray}
The new feature of this type of model is that this parameter need not be 
small in which case some new features in the generation of neutrino masses and 
mixing can result.
 
We postulate the existence of the following fields and associated $U(1)_H$
charges
\begin{eqnarray}
L_{+1}, L_0, L_0, N_{+1}, \bar{N}_{-1}, N_0, \bar{N}_0 \;.
\label{def}
\end{eqnarray}
The fields $L$ are the three generations of lepton doublets in the Standard
Model, while the fields $N$ are the Standard Model singlets. The fields
$L$ and $\bar{N}$ give rise to the usual Dirac mass terms. Vector-like 
couplings can arise between $N$ and $\bar{N}$ while lepton violating couplings
arise between two $N$ fields or two $\bar{N}$ fields. 
Setting the $U(1)_H$ charges equal in the lepton doublet fields can 
yield large mixing angles in the neutrino mass matrix and has been 
called neutrino anarchy\cite{Hall:1999sn,Berger:2000sc,Hirsch:2001mw,Vissani:2001im,Espinosa:2003qz}.
The charges in Eq.~(\ref{def}) guarantee a semi-anarchy\cite{Altarelli:2004za}
in the second and third generations 
(because the $U(1)_H$ charges are the same), and will
also exhibit an inverted hierarchy. 
This collection of heavy neutrino fields is quite economical. Other fields can 
be present without changing the features of the light neutrino mass matrix.
As mentioned in the introduction 
a new ingredient we introduce here is to assign a nonzero horizontal charge 
to the lepton number violating scalar fields $S_{HL}$ and 
$\bar{S}_{HL}$. First consider the situation where the horizontal symmetry 
breaking is turned off (i.e. $\lambda_H=0$).
With the charges of the fields chosen as in Eq.~(\ref{def}), the
following symmetric mass matrix results:
\begin{eqnarray}
\left (\begin{array}{ccccccc}
0 & 0 & 0 & 0 & \phi_u & 0 & 0 \\ 
0 & 0 & 0 & 0 & 0 & 0 & \phi_u \\
0 & 0 & 0 & 0 & 0 & 0 & \phi_u \\
0 & 0 & 0 & 0 & M_F & S_{HL} & 0 \\
\phi_u & 0 & 0 & M_F & 0 & 0 & \bar{S}_{HL} \\
0 & 0 & 0 & S_{HL} & 0 & 0 & M_F \\
0 & \phi_u & \phi_u & 0 & \bar{S}_{HL} & M_F & 0  
\end{array}\right )\;.
\label{modelu}
\end{eqnarray}
In this matrix the entries $S_{HL}$ and $\bar{S}_{HL}$ represent their 
vevs, $M_{HL}=\langle S_{HL}\rangle \sim \langle \bar{S}_{HL}\rangle$.
We imagine the scale $M_{HL}$ is comparable to the scale $M_F$ so that the 
superheavy $4\times 4$ subblock is fully mixed. This 
implies that the $\lambda_L$ parameter in Eq.~(\ref{lambdal}) should be 
sufficiently small, $\lambda_L\alt \lambda _H)$. This is the only condition 
on $\lambda_L$ and our results can henceforth be described in terms of the 
small parameter $\lambda _H$ alone.
The effect of assigning a nonzero $U(1)_H$ charge to the fields giving
rise to lepton number
violation, $S_{HL}$ and $\bar{S}_{HL}$, 
is to move the contributions off the diagonal, and
as a consequence generate a pseudo-Dirac structure in the effective 
light neutrino mass matrix.
It should be understood that there are undetermined order one coefficients
in front of every nonzero term in the matrix as in all model of this 
Froggatt-Nielsen type. Some of the entries are 
related because the matrix is symmetric, but other entries which appear 
identical are only equivalent up to these coefficients (for example, $M_F$'s 
in the $4-5$ entry and in the $7-8$ entry).

Integrating out the heavy sector gives an effective light neutrino mass matrix
\begin{eqnarray}
m\left (\begin{array}{ccc}
0 & \sin \theta & \cos \theta \\
\sin \theta & 0 & 0 \\
\cos \theta & 0 & 0 
\end{array}\right )\;.
\label{unperturb}
\end{eqnarray}
The overall scale $m$ is generated as a seesaw-type mass 
of order ${\cal O}(\phi_u^2/S_{HL})$ and the 
undetermined ratio of order one coefficients has been expressed as an 
angle.
This matrix is of much interest because it 
has two large angles and a small one. It can be understood as 
involving large mixing 
between the second and third family, together with a pseudo-Dirac structure
relating the first family to them. The pseudo-Dirac case is 
characterized in the two generation
case by the matrix
\begin{eqnarray}
m\left (\begin{array}{ccc}
0 & 1  \\
1 & 0 
\end{array}\right )\;.
\label{pseudoDirac}
\end{eqnarray}
and the matrix in Eq.~(\ref{unperturb}) is clearly of this form. 

The effective light neutrino mass matrix in Eq.~(\ref{unperturb}) 
is a well-known mass matrix form that almost gives acceptable values for the 
neutrino parameters
\begin{eqnarray}
\sin ^2 2\theta _A&=&\sin^2 2\theta\;, \nonumber \\
\sin ^2 2\theta_\odot &=& 1\;,
\end{eqnarray}
where the $A$ and $\odot$ subscripts refer to the atmospheric and solar angles
respectively. The mass eigenvalues of the matrix are
$0$, $-m$, and $+m$. 
As noted earlier this
structure is an interesting starting point for constructing realistic 
masses and mixing angles to describe the neutrino oscillation data. 
The atmospheric neutrino mixing angle is the same as $\theta$ which can be 
large. In our formulation 
this angle is anarchical because the muon and tau lepton 
doublet horizontal charges are equal to each other. The other large angle 
is in fact maximal, and may be large for a fundamentally different reason. 
We will
exploit this distinction between the two large angles in our 
model\footnote{Sometimes an additional symmetry is imposed to force 
$\sin \theta = \cos \theta= 1/\sqrt{2}$ so the the atmospheric neutrino 
mixing angle is maximal. This gives a bimaximal neutrino mixing scenario, but
as we have argued here, the maximality of the solar and atmospheric neutrino 
angles then arises in different and generically unrelated ways.}. 

The mixing angles observed in neutrino oscillation experiments are the angles 
in the matrix
\begin{eqnarray}
U_{\rm PMNS}=U_L^\dagger U_\nu\;,
\end{eqnarray}
where $U_L$ is the matrix that diagonalizes the charged lepton sector, while
$U_\nu$ diagonalizes the effective light neutrino mass matrix. It is
expected in Froggatt-Nielsen models of the type we are considering that the 
mixing angles in the charged lepton sector should be small and CKM-like since 
the electron, muon and tau lepton exhibit a strong hierarchy in masses. 
Therefore we expect statements about the mixing angles arising from $U_\nu$ 
to provide the substantial contributions to the mixing angles observed in the 
experiments in $U_{\rm PMNS}$.

The mixing matrix $U_\nu$ has one maximal angle, one angle $\theta$
which is given in terms of (undetermined) 
order one coefficients in the full mass matrix in Eq.~(\ref{modelu}). 
In fact, because of the symmetries of the model,
the source of the light neutrino
masses arises from a limited number of interactions, and 
$\tan \theta$ is simply the ratio of the $3-7$ and the $2-7$ 
entries in the matrix in Eq.~(\ref{modelu}). 
(For the special case $\tan \theta=1$ one has 
maximal mixing for the atmospheric neutrinos.)
The remaining angle is such that $U_{e3}=0$.

With a small perturbation the mass matrix in Eq.~(\ref{unperturb})
can be made roughly consistent with the experimental data which requires 
a solar mixing angle $\theta_\odot \approx 35^o$ somewhat smaller than maximal
and a mass splitting between the two nonzero eigenvalues to account for 
a small $\Delta m_\odot^2$. 

Including the effects of the field $S_H$ (through the coupling 
$S_HN_{+1}\bar{N}_0$ which conserves the horizontal charge $U(1)_H$)
gives a contribution that can be 
regarded as a perturbation
\begin{eqnarray}
\left (\begin{array}{ccccccc}
0 & 0 & 0 & 0 & \phi_u & 0 & 0 \\ 
0 & 0 & 0 & 0 & 0 & 0 & \phi_u \\
0 & 0 & 0 & 0 & 0 & 0 & \phi_u \\
0 & 0 & 0 & 0 & M_F & S_{HL} & PM_F \\
\phi_u & 0 & 0 & M_F & 0 & 0 & \bar{S}_{HL} \\
0 & 0 & 0 & S_{HL} & 0 & 0 & M_F \\
0 & \phi_u & \phi_u & PM_F & \bar{S}_{HL} & M_F & 0  
\end{array}\right )\;.
\label{model}
\end{eqnarray}
where $P={\cal O}(M_H/M_F)={\cal O}(\lambda_H)$. This perturbs the matrix in 
Eq.~(\ref{unperturb}) to the following form,
\begin{eqnarray}
m\left (\begin{array}{ccc}
z & \sin \theta & \cos \theta \\
\sin \theta & 0 & 0 \\
\cos \theta & 0 & 0
\end{array}\right )\;,
\label{perturb}
\end{eqnarray}
where the small quantity $z$ is of order the perturbation,
$P$.

The lepton numbers can be assigned to the fermion fields as in 
Table~\ref{lepton}. These are consistent with the $U(1)_L$ symmetry which 
ensures lepton number $L=L_e+L_\mu+L_\tau$ conservation in that all the 
interactions conserve overall lepton number $L$.

The underlying $L_e-L_\mu-L_\tau$ symmetry can be understood from the 
charges in Eq.~(\ref{def}).
Charges can be assigned to the fields $S_{HL}$ and $\bar{S}_{HL}$ so that 
the interactions generating the masses from 
$M_{HL}$ and the Dirac masses $\phi$ all respect this symmetry. The 
field $S_H$ whose usual role is to provide the small parameter that gives
rise to flavor hierarchies in the Standard Model, here also 
provides a source for the breaking of the $L_e-L_\mu-L_\tau$ symmetry. This
symmetry 
arises as an accidental symmetry due
to the charge assignments of the fields, 
and does not necessarily arise for some more 
fundamental reason. Nevertheless it 
would be interesting to see if this scheme could 
arise from a grand unified model.

\begin{table}[t]
\caption{\label{Lepton charges.}}
\begin{tabular}{|c|r|r|r|r|r|}
\hline
       & $L_e$ & $L_\mu + L_\tau$ & $L_e-L_\mu -L_\tau$ & $L$ & $H$ \\ 
\hline
$L_{+1}$ &   $+1$   &        $0$         &      $+1$           & $+1$ & $+1$\\ 
$L_0$    &   $0$   &        $+1$         &      $-1$           & $+1$ & $0$ \\ 
$L_0$    &   $0$   &        $+1$         &      $-1$           & $+1$ & $0$ \\ 
$N_{+1}$ &   $+1$   &       $0$         &       $+1$           & $+1$ & $+1$\\ 
$\bar{N}_{-1}$ & $-1$ &      $0$         &      $-1$           & $-1$ & $-1$\\
$N_{0}$ &   $0$   &         $+1$         &      $-1$           & $+1$ & $0$ \\ 
$\bar{N}_{0}$ & $0$ &       $-1$         &      $+1$           & $-1$ & $0$ \\
\hline
\end{tabular}
\label{lepton}
\end{table}

The $S_H$ and $S_H^\prime$ fields are singlets 
under $U(1)_L$ while the fields $S_{HL}$ and 
$\bar{S}_{HL}$ are charged. Since the $S_H$ field is presumably responsible
for the hierarchy in the quark and charged lepton sectors, we assign it 
zero also under the $L_e-L_\mu-L_\tau$ symmetry\footnote{Alternatively
one can assign individual lepton number charges to the $S_H$ and $S_H^\prime$ 
fields in which case the $L_e-L_\mu-L_\tau$ symmetry only arises after they 
acquire vevs. However then the fields do not respect the $U(1)_L$ symmetry
in the interactions with the quark sector.}. These fields are assigned the 
$L_e-L_\mu-L_\tau$ charges according  
to Table~\ref{scalar}.

\begin{table}[t]
\caption{\label{Lepton charges.}}
\begin{tabular}{|c|r|r|r|r|r|}
\hline
       & $L_e$ & $L_\mu + L_\tau$ & $L_e-L_\mu -L_\tau$ & $L$ & $H$\\ 
\hline
$S_{HL}$    &   $-1$   &        $-1$         &      $0$       & $-2$ & $-1$\\ 
$\bar{S}_{HL}$    &   $+1$   &        $+1$         &      $0$ & $+2$ & $+1$\\ 
$S_H$    &      $0$   &        $0$         &      $0$      & $0$ & $-1$ \\ 
$S_H^\prime$ &  $0$   &        $0$         &      $0$      & $0$ & $+1$ \\ 
\hline
\end{tabular}
\label{scalar}
\end{table}

The interaction $S_HN_{+1}\bar{N}_0$ has charge $L_e-L_\mu -L_\tau=+2$, and 
so does not respect the $L_e-L_\mu -L_\tau$ symmetry and generates mass terms
in the full neutrino matrix when $S_H$ gets 
a vev. It 
fills in the $1-1$ entry of the light neutrino mass matrix involving the 
parameter $z$. If there is a new field $S_H^\prime$ 
with $U(1)_H$ charge $+1$, then it provides a coupling
$S_H^\prime N_{0}\bar{N}_{-1}$ has charge $L_e-L_\mu -L_\tau=-2$. This
interaction generates a perturbation for the 
$2-3$ subblock in the light neutrino mass matrix.

\begin{eqnarray}
\left (\begin{array}{ccccccc}
0 & 0 & 0 & 0 & \phi_u & 0 & 0 \\ 
0 & 0 & 0 & 0 & 0 & 0 & \phi_u \\
0 & 0 & 0 & 0 & 0 & 0 & \phi_u \\
0 & 0 & 0 & 0 & M_F & S_{HL} & 0 \\
\phi_u & 0 & 0 & M_F & 0 & P^\prime M_F & \bar{S}_{HL} \\
0 & 0 & 0 & S_{HL} & P^\prime M_F & 0 & M_F \\
0 & \phi_u & \phi_u & 0 & \bar{S}_{HL} & M_F & 0  
\end{array}\right )\;.
\label{model2}
\end{eqnarray}
where $P^\prime={\cal O}(\lambda_H)$. The effective light neutrino mass matrix in 
Eq.~(\ref{unperturb}) becomes,
\begin{eqnarray}
m\left (\begin{array}{ccc}
0 & \sin \theta & \cos \theta \\
\sin \theta & z'\sin ^2\theta & z'\sin \theta\cos \theta \\
\cos \theta & z'\sin \theta \cos \theta & z'\cos ^2\theta 
\end{array}\right )\;,
\label{perturb2}
\end{eqnarray}
where the small quantity $z^\prime$ is of order the perturbation,
$P^\prime$. The relationship between the four terms in the $2-3$ subblock
of this matrix is enforced by the restricted nature of the coupling of the 
lepton doublets $L$ to the heavy neutrino fields $N$ and $\bar{N}$.

To obtain a good fit to the solar and atmospheric neutrino data we will 
need to include the effects of both types 
of $L_e-L_\mu-L_\tau $ symmetry breaking (see the next section).
Including both perturbations $P$ and $P^\prime$,
\begin{eqnarray}
\left (\begin{array}{ccccccc}
0 & 0 & 0 & 0 & \phi_u & 0 & 0 \\ 
0 & 0 & 0 & 0 & 0 & 0 & \phi_u \\
0 & 0 & 0 & 0 & 0 & 0 & \phi_u \\
0 & 0 & 0 & 0 & M_F & S_{HL} & PM_F \\
\phi_u & 0 & 0 & M_F & 0 & P^\prime M_F & \bar{S}_{HL} \\
0 & 0 & 0 & S_{HL} & P^\prime M_F & 0 & M_F \\
0 & \phi_u & \phi_u & PM_F & \bar{S}_{HL} & M_F & 0  
\end{array}\right )\;,
\label{model3}
\end{eqnarray}
yields
a light neutrino mass matrix that can be phenomenologically acceptable for
describing all the data from atmospheric and solar neutrino oscillations,
\begin{eqnarray}
m\left (\begin{array}{ccc}
z & \sin \theta & \cos \theta \\
\sin \theta & z'\sin ^2\theta & z'\sin \theta\cos \theta \\
\cos \theta & z'\sin \theta \cos \theta & z'\cos ^2\theta 
\end{array}\right )\;.
\label{perturb3}
\end{eqnarray}
This matrix has one massless neutrino and $U_{e3}=0$.

\section{Physical predictions of the model}

Neutrino mass matrices that are pertubations of the form in 
Eq.~(\ref{unperturb}) have been studied in a more general context 
in Ref.\cite{Goh:2002nk}. There a perturbation is
added so that the symmetric matrix 
\begin{eqnarray}
m\left (\begin{array}{ccc}
z & \sin \theta & \cos \theta \\
\sin \theta & y & d \\
\cos \theta & d & x 
\end{array}\right )\;,
\label{Goh}
\end{eqnarray}
results. Our light neutrino mass matrices in 
Eqs.~(\ref{perturb}), (\ref{perturb2}) and (\ref{perturb3}) 
are examples of this more 
general form. 
This mass matrix involving perturbations $\delta$ of order $x,y,z,d$
gives\cite{Goh:2002nk} 
\begin{eqnarray}
\sin^22\theta_{A}&=&\sin^22\theta + {\cal O}(\delta^2)\;, \nonumber \\
\triangle m_A^2&=&-m^2 + 2 \triangle m^2_{\odot}+ {\cal O}(\delta^2)\;,
\label{first}
\end{eqnarray}
and
\begin{eqnarray}
\sin^22\theta_{\odot}~=~1-\left (\frac{\triangle m_\odot^2}{4\triangle
m_A^2}-z\right )^2~+~{\cal O}(\delta^3) \cr
R\equiv \frac{\triangle m_\odot^2}{\triangle
m_A^2}~=~2(z+\vec{v}\cdot\vec{x})~+~{\cal O}(\delta^2)\cr
U_{e3}~=~\vec{A}\cdot(\vec{v}\times\vec{x})~+~{\cal O}(\delta^3)
\label{second}
\end{eqnarray} 
where $\vec{v}=(\cos^2\theta,\sin^2\theta,\sqrt{2}\sin\theta\cos\theta)$, 
$\vec{x}=(x,y,\sqrt{2}d)$ and
$\vec{A}~=~\frac{1}{\sqrt{2}}(1,1,0)$. 

For the light neutrino mass matrix in Eq.~(\ref{perturb3}) the following 
values for the ratio of mass-squares and solar mixing angle result,
\begin{eqnarray}
\sin^2 2\theta_\odot&=&1-{1\over 4}(z-z^\prime)^2\;, \nonumber \\
R&=&2(z+z^\prime)\;, \nonumber \\
U_{e3}&=&0\;. 
\label{third}
\end{eqnarray}
This last relation is in fact exact.
If one of the perturbations $P$ or $P^\prime$ is turned off, then it is 
clear that one obtains the phenomenologically unacceptable relationship
\begin{eqnarray}
\sin^2 2\theta_\odot&=&1-{R^2\over {16}}\;.
\end{eqnarray}
The small observed value of $R\sim 0.04$ and the deviation
of the solar mixing angle, $\theta_\odot\sim 35^o$, from maximal is a source
of tension in all models employing a perturbative breaking of the 
$L_e-L_\mu-L_\tau$ symmetry involving a single parameter. 

Given this general analysis of the matrix in Eq.~(\ref{Goh}), we can make the 
following conclusions about our model:
The ratio of the mass-squared parameters satisfies 
$\Delta m_\odot^2/\Delta m_A^2\sim P,P^\prime$. 
The atmospheric neutrino mixing angle
is large as a result of anarchy and satisfies the first equation in 
Eq.~(\ref{first}) with no ${\cal O}(\delta^2)$ corrections. 

The solar neutrino mixing angle satisfies 
$\sin^22\theta_{\odot}=1-{\cal O}(P^2,P^{\prime 2})$. 
Since $v$ is parallel to $x$ we have
$U_{e3}=0$. Finally the neutrinoless double beta decay parameter in the 
general case of Eq.~(\ref{Goh}) is given 
by\cite{Goh:2002nk} 
$\langle m\rangle _{\beta\beta}=zm$. In our model this parameter 
is therefore set to the scale 
$\sim \lambda_H\sqrt{\Delta m_A^2}$.

The dependence of the solar mixing angle and the ratio of the mass-squares in 
Eq.~(\ref{third}) indicates that a small fine-tuning $z\approx -z^\prime$ can
give rise to acceptable values, $1-\sin^2 2\theta_\odot\sim \lambda_H^2$ and 
$R\sim \lambda _H^2$. The terms which break the $L_e-L_\mu-L_\tau$
symmetry do so in such a way to generate entries in a very specific form in 
the light neutrino mass matrix. In fact this simple structure gives a 
prediction for the mixing angle, $U_{e3}=0$.

It is easy to adjust the charge assignments to make the perturbations $P$ and
$P^\prime$ any 
power of the parameter $\lambda _H$. We have chosen 
$P,P^\prime \sim \lambda _H$ because
this gives the best overall fit to the experimental data. 

\section{Summary and Conclusions}

The model we have constructed in this note exhibits the following features:

$\bullet$ The large mixing in the atmospheric neutrino oscillations results by 
assuming the same horizontal $U(1)_H$ charge for the second and third 
flavor generation for the $SU(2)_L$ lepton doublets of the Standard Model. 
This is the usual anarchy in the neutrino sector in Froggatt-Nielsen models,
and is sometimes called a semi-anarchy since it does not extend to the first
generation.

$\bullet$ The large mixing in the solar neutrino oscillations results from 
the fact that there are two distinct kinds of couplings for the heavy neutrino
sector: 1) there are lepton-number violating Majorana masses of order $M_{HL}$,
and 2) there are vector-like mass of order $M_F$. The resulting
large angle in the 
light neutrino mass matrix becomes the solar neutrino mixing angle and
is of the 
pseudo-Dirac form. The feature of the model which resulted in this 
pseudo-Dirac mass matrix is the nonzero $U(1)_H$ charge
of the fields $S_{HL}$ and $\bar{S}_{HL}$ that give rise to lepton 
number violation.

$\bullet$ In Froggatt-Nielsen models it is difficult to achieve a light 
neutrino mass matrix that has two large mixing angles and a small third
mixing ($U_{e3}$). In the model presented in this paper the source
of the two large angles is separated into the distinct mechanisms of the 
previous bullet items. 

$\bullet$ Perturbations are present in the model in terms of the 
usual Froggatt-Nielsen parameter $\lambda _H$ that can yield
a realistic set of mass-squares and mixing angles.
The model has an inverted hierarchy with one massless neutrino, and it 
requires $U_{e3}=0$ barring small contributions from the charged lepton 
sector. The solar neutrino mixing angle satisfies 
$1-\sin^22\theta_\odot\sim P^2,P^{\prime 2}$. 
The ratio of mass-squared parameters for the neutrino oscillation experiments
satisfy $R=\Delta m_\odot^2/\Delta m_A^2\sim P,P^\prime$ where $P$ and 
$P^\prime$ are perturbations 
related to the small horizontal symmetry paratemer (Cabibbo angle) of 
$\lambda _H\sim 0.2$. A modest amount of fine tuning between the two sources
of perturbation are needed to suppress the ratio $R$ to an acceptable value.

In summary we have constructed a model which ``combines'' 
lepton number violation
with the horizontal symmetry violation. By assigning the lepton number 
violating field $S_{HL}$ a horizontal symmetry charge, a pseudo-Dirac 
structure is imposed on the light neutrino mass matrix. Since the model 
contains two scales associated with the breaking of lepton number on the 
one hand and the breaking of the horizontal symmetry on the other, the 
small Froggatt-Nielsen parameter naturally introduces the needed perturbations
to the $L_e-L_\mu-L_\tau $ symmetry to accommodate the experimental data.
This idea may be of more general use in flavor models 
based on the Froggatt-Nielsen mechanism. 

\section*{Acknowledgments}
This work was supported in part by the U.S.
Department of Energy under Grant No.~DE-FG02-91ER40661.

{\it Note added}: 
After this work was submitted to the arXiv, we learned that our
model falls into a class of models governed by the scaling ansatz introduced
in Ref.~\cite{Mohapatra:2006xy}. Some further description of the phenomenology
of these models can be found there.

\end{document}